\documentstyle[11pt,newpasp,twoside,psfig]{article}

\begin{document}

\title{Nanoarcsecond Single-Dish Imaging of the Vela 
Pulsar}

\author{J.-P. Macquart, S. Johnston, M. Walker}
\affil{Special Research Centre for Theoretical Astrophysics, School 
of Physics, University of Sydney, New South Wales 2006, Australia}

\author{D. Stinebring}
\affil{Department of Physics, Oberlin College, Oberlin, OH 44074}

\begin{abstract}
We have measured the properties of the diffractive scintillation 
toward the Vela pulsar under the extremely strong scattering 
conditions encountered at 660~MHz.  We obtain a 
decorrelation bandwidth of $\nu_d = 244 \pm 4$~Hz and diffractive 
decorrelation timescale of $t_{\rm diff} = 3.3\pm 0.3$~s.  Our 
measurement of the modulation indices $m=0.87\pm 0.003\pm 0.05$ and 
$m=0.93\pm 0.03 \pm 0.05$ (one for each polarization stream), are at 
variance with the 
modulation index of the Vela pulsar obtained at 2.3~GHz by 
Gwinn et al. (1997) {\it if} the deviation from a modulation index of 
unity is ascribed to a source size effect.  
\end{abstract}

\section{Introduction}
Under sufficiently strong scattering conditions, a point source whose 
radiation propagates through the Interstellar Medium (ISM) 
exhibits diffractive scintillation, whereby the fluctuations in the 
intensity, $I$, are fully modulated: $\langle (I - \langle I \rangle)^2 
\rangle/ \langle I \rangle^2 = 1$.  If $r_{\rm diff}$ is the 
length scale on the scattering screen for which the root-mean-square phase 
difference is one radian, the angular size of the diffractive 
pattern is $\theta_d = r_{\rm diff}/D$, where $D$ is distance  
between the observer and the scattering screen.   The amplitude of the intensity 
fluctuations is suppressed if the angular size of a scintillating 
object, $\theta_s$, is comparable to $\theta_d$.   

Finite source size effects are more pronounced at low frequency. 
Since the phase delay caused by the density inhomogeneities in the ISM is linearly 
proportional to frequency, the scatttering effect 
of density inhomogeneities is greater at lower frequency.  The diffractive 
scale, which is a measure of the amplitude of the 
turbulent phase fluctuations, therefore 
decreases with frequency.  Pulsar scattering data shows that the 
spectrum of turbulent density fluctuations in the ISM, $\Phi({\bf q})$, is 
consistent with a power law: $\Phi({\bf q}) \propto q^{-\beta}$ over 
$\sim 5$ decades in wavenumber (Armstrong, Rickett \& Spangler 1995).
For such behaviour, the diffractive scale varies as 
$r_{\rm diff} \propto \nu^{2/(\beta-2)}$, with evidence for both 
$\beta \approx 11/3$ and  $\beta\approx 4$ for the Vela pulsar (Johnston et 
al. 1998).  
 
Gwinn et al. (1997) (see also these proceedings) measured $m=0.87$ 
for the Vela pulsar at 2.3~GHz and, attributing the deviation from 
$m=1$ to a source size effect,  
used the theory of diffractive scintillation to derive a source size 
of $\sim 500$~km.  However, as $\theta_d/\theta_s$ decreases 
for the stronger scattering encountered at lower frequencies the 
modulation index is also expected to decrease.  For the source 
size stated by Gwinn et al. (1997), scintillation theory predicts that 
the modulation index is no larger than 0.45 if $\beta=4$ ($m<0.35$ 
if $\beta=11/3$) at 660~MHz, assuming that the size of the 
emission region does not decrease with frequency.

Conversely, source-size effects at higher frequencies are expected to 
be negligible, with the intensity probability distribution following 
a negative exponential distribution $p(I) = 1/I_0 \exp(-I/I_0)$ with 
mean intensity $I_0$ (e.g. 
Gwinn et al. 1998).

\section{Results}
We observed the Vela pulsar for 3~minutes with the Parkes telescope and the CPSR 
(Caltech-Parkes-Swinburne Recorder) backend to analyse the 
diffractive scintillation of Vela at 660~MHz, thereby testing the source-size 
assertion of Gwinn et al. (1997).   
We describe the data reduction procedure here briefly; full details of 
the analysis will be presented elsewhere (Macquart et al. 2000).

The CPSR system recorded a two-bit complex sampled data stream in each of two linear 
polarizations at a rate of 20~MHz.  Each polarization stream was 
analysed separately.  For each stream, the mean pulsar 
power was determined by subtracting the average off-pulse spectrum (obtained by 
FFTing the data stream) from the on-pulse spectrum.  Although the 
variation of the pulsar flux density is negligible over the 20~MHz 
bandwidth, Faraday rotation across the band is not, as it causes the 
detected power in each (linear) polarization stream to vary as a function of 
frequency.

The spectra then were combined in groups of 10 pulses -- 
equivalent to a third of the scintillation timescale -- 
and normalized by the pulsar's mean power at that frequency and by 
the instrumental bandpass.  The mean signal across the normalised band 
was subtracted to leave only the fluctuations in $I(\nu)$ across the 
band.  The outer eighths of the band were clipped due to the tapering of 
the bandpass at the edges.  These normalised pulsar spectra were then 
autocorrelated and cross-correlated to find the normalised covariance
\begin{eqnarray}
    \Gamma(\Delta \nu,\Delta t) = \frac{\langle[I(\nu+\Delta 
    \nu,t+\Delta t)-\langle I(\nu,t) \rangle]^2 \rangle}{
    \langle I(\nu,t) \rangle^2}.
\end{eqnarray}

\begin{figure*}[h]
\centering
\begin{tabular}{c}
\psfig{file=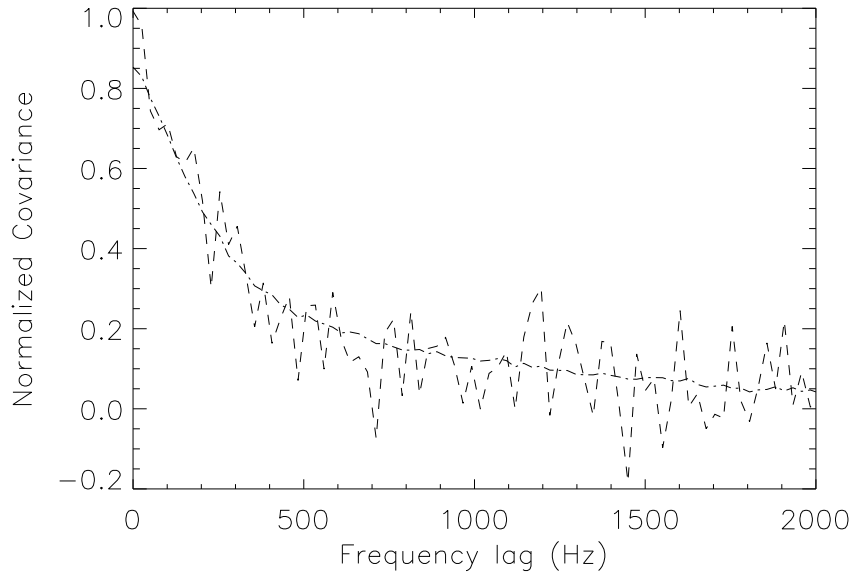,width=120mm}
\end{tabular}

\caption{Spectral decorrelation of the signal in 
the two polarization channels.  The zero-lag spikes due to receiver noise 
are omitted.  The dot-dash line represents poln~0 data, the dashed 
line poln~1 data.  Note the difference in signal-to-noise between the two 
polarization streams.  The high ($\sim 80$\%) linear 
polarization of the Vela pulsar at this frequency and the parallactic 
angle during the observation combined to ensure that little power was 
collected in the poln~1.  Since the signal-to-noise is lower in this 
polarization, the noise in the autocorrelation function is correspondingly 
greater.}
\end{figure*}

Figure 1 shows the frequency autocorrelation function $\Gamma(\Delta 
\nu,0)$.  Fits to the covariance function 
yielded a decorrelation bandwidth $\nu_d = 244 \pm 4$~Hz for poln~0 
($\nu_d = 241 \pm 8$~Hz for poln 1) and a decorrelation timescale 
$t_{\rm diff} = 3.3 \pm 0.3$~s determined from poln~0 
only due to the higher signal to noise available in this channel.  The 
modulation indices are $m=0.871\pm 0.003$ (poln~0) and $m=0.93 \pm 
0.03$ (poln 1).  The stated errors are obtained from those formally 
obtained from the data shown in figure~1.  However, these measurements 
are subject to other errors:
\begin{itemize}    
\item {\it Intrinsic pulse-to-pulse flux variations} may affect 
the scaling of the modulation indices.  Suppose we receive 
a pulse $Y$ times stronger than the mean pulse flux density.  Then the 
scintillation signal, which was normalised by the mean pulse flux 
density over one minute, is 
thus measured to have a modulation index $m_{\rm meas} = Y m_{\rm 
real}$.  The effect of intrinsic pulse 
variability is reduced by averaging together many independent pulses 
before calculating the autocorrelation function, and is further 
reduced by averaging together many such autocorrelation functions.  Over the 
3~min of data used, intrinsic pulse variations contribute an 
error $\Delta m \approx 0.01$.

\item {\it Telescope gain variations} play a similar r\^ole to intrinsic pulsar 
variability.  We estimate their effect on timescales of one minute to be less than a few percent.  

\item The {\it finite spectral and temporal resolution} of our observation may 
{\it reduce} our measurement of the modulation index due to smearing 
of the scintillation pattern.   However, 
the spectral and temporal resolution used are sufficiently small compared 
to the decorrelation bandwidth and timescale that these effects are 
negligible. 
\end{itemize}

In total, we estimate that the total error in our measured modulation 
index due to the effects mentioned above is no more than 5\%.

In addition to the 660~MHz data, we have also obtained scintillation 
data at a frequency of 8.4~GHz (Macquart et al. 2000), for which one 
expects $m=0.99$ (i.e. negligible source-size effects).  However, the 
observed intensity distribution deviates significantly below the 
expected negative exponential distribution at high intensity, and the 
modulation index is $m \approx 0.93$.

\section{Discussion}
Our measured modulation index $m=0.87 \pm 0.05$ is significantly at 
variance with the modulation index of $m<0.45$ expected if the 
quenching of the diffractive scintillation at 2.3~GHz is a 
source-size effect.  The 660~MHz scintillation data places an upper 
limit of 50~km on the size of this region.

The two main explanations for the apparent contradiction are: 
\begin{itemize}
\item The pulsar's emission region is {\it smaller} at low frequency.  
The upper limit on the expected modulation index assumes 
that the emission region retains the same characteristic size between 
2.3~GHz and 660~MHz; however, the radius to frequency mapping paradigm 
leads one to expect the emission region to be larger at low 
frequency and thus that the diffractive scintillation is quenched even 
further.  
\item The reduction in $m$ observed by Gwinn et al. is not related to the 
apparent source size.  This explanation is supported by the 8.4~GHz 
scintillation data.  This data may indicate that the statistics of 
the phase fluctuations on the scattering screen are not Gaussian, and 
thus that the expected intensity distribution due to scintillation 
of a point source is not exactly negative exponential in form.
\end{itemize}

For a complete discussion of the observing and calibration procedures 
used and further implications of the results see Macquart et al. (2000).

\end{document}